\RequirePackage{xcolor}
\documentclass[journal,twoside,web]{ieeecolor}

\usepackage{etoolbox}
\makeatletter
\@ifundefined{color@begingroup}%
  {\let\color@begingroup\relax
   \let\color@endgroup\relax}{}%
\def\fix@ieeecolor@hbox#1{%
  \hbox{\color@begingroup#1\color@endgroup}}
\patchcmd\@makecaption{\hbox}{\fix@ieeecolor@hbox}{}{\FAILED}
\patchcmd\@makecaption{\hbox}{\fix@ieeecolor@hbox}{}{\FAILED}

\usepackage{tmi}
\usepackage{graphicx}
\usepackage{textcomp}
\usepackage{booktabs}
\usepackage{float}
\usepackage{physics}
\usepackage{amsmath}
\usepackage{amssymb}
\usepackage{amsfonts}
\usepackage{cite}
\usepackage{multirow}
\usepackage{tabularx}
\usepackage[hidelinks]{hyperref}
\usepackage{placeins}
\usepackage{threeparttable}
\usepackage[referable]{threeparttablex}

\def\BibTeX{{\rm B\kern-.05em{\sc i\kern-.025em b}\kern-.08em
    T\kern-.1667em\lower.7ex\hbox{E}\kern-.125emX}}
\begin{document}
\title{Comprehensive Analysis of Relative Pressure Estimation in Type B Aortic Dissection Utilizing 4D-Flow MRI}
\author{Brandon Hardy, Judith Zimmermann, Vincent Lechner, Mia Bonini, Julio A. Sotelo, Nicholas S. Burris, Daniel B. Ennis, David Marlevi, and David A. Nordsletten \\
    \vspace{5mm} \textit{Dedicated to the memory of Brian Rupert, MD, who passed suddenly from aortic dissection}
    \thanks{\copyright~2026 IEEE. Personal use of this material is permitted. Permission from IEEE must be obtained for all other uses, in any current or future media, including reprinting/republishing this material for advertising or promotional purposes, creating new collective works, for resale or redistribution to servers or lists, or reuse of any copyrighted component of this work in other works.}
    \thanks{Submitted for review on 05/11/2025. This work was supported by the National Institutes of Health under R01 HL183592 to DBE and R01 HL170059 to DAN\@. The views expressed are those of the authors and not necessarily those of the NIH.}
    \thanks{Brandon Hardy is with the Department of Biomedical Engineering, University of Michigan, Ann Arbor, USA (e-mail: bkhardy@umich.edu).}
    \thanks{Judith Zimmermann is with the Department of Radiology, Stanford University, Stanford, USA (e-mail: juzim@stanford.edu).}
    \thanks{Vincent Lechner is with the Department of Molecular Medicine and Surgery, Karolinska Institutet, Stockholm, SWE (e-mail: vincent.lechner@ki.se).}
    \thanks{Mia Bonini is with the Department of Biomedical Engineering, University of Michigan, Ann Arbor, USA (e-mail: mbonini@umich.edu).}
    \thanks{Julio A. Sotelo is with the Departamento de Informática, Universidad Técnica Federico Santa María, Santiago, CHL (e-mail: julio.sotelo@usm.cl).}
    \thanks{Nicholas S. Burris is with the Department of Radiology, University of Michigan, Ann Arbor, USA and the Department of Radiology, University of Wisconsin--Madison, Madison, USA (e-mail: nsburris@wisc.edu).}
    \thanks{Daniel B. Ennis is with the Department of Radiology, Stanford University, Stanford, USA (e-mail: dbe@stanford.edu).}
    \thanks{David Marlevi is with the Department of Molecular Medicine and Surgery, Karolinska Institutet, Stockholm, SWE and the Institute for Medical Engineering \& Science, Massachusetts Institute of Technology, Cambridge, USA (e-mail: marlev@ki.se).}
    \thanks{David Nordsletten is with the Department of Biomedical Engineering and the Department of Cardiac Surgery, University of Michigan, Ann Arbor, USA, and the School of Biomedical Engineering and Imaging Sciences, King's College London, UK (e-mail: nordslet@umich.edu).}
}

\maketitle

\begin{abstract}
    Magnetic resonance imaging (MRI) can estimate three-dimensional (3D) time-resolved blood velocity fields using 4D-flow MRI, providing rich relative pressure field information.
    Clinical alternatives---catheterization and Doppler echocardiography---provide one-dimensional pressure drops.
    While the accuracy of one-dimensional pressure drops from 4D-flow has been explored, the accuracy of 3D relative pressure field estimates requires further evaluation.
    This work analyzes three state-of-the-art relative pressure estimators: virtual Work-Energy Relative Pressure (\textit{v}WERP), the Pressure Poisson Estimator (PPE), and the Stokes Estimator (STE).
    Spatiotemporal characteristics and noise sensitivity were determined \textit{in silico} via comparison with CFD-derived relative pressure fields.
    Estimators were then validated using a type B aortic dissection (TBAD) flow phantom with varying tear geometry and twelve catheter pressure measurements.
    Finally, each estimator was evaluated across eight patient cases in a clinical feasibility analysis. Estimators were cross-compared and evaluated for coherence since no reference catheter pressure measurements were available.
    PPE outperformed STE in the \textit{in silico} relative pressure analysis, while both outperformed \textit{v}WERP in plane-to-plane assessment.
    High velocity gradients and low spatial resolution contributed most significantly to local variations in 3D relative pressure field errors.
    Low temporal resolution led to systematic underestimation of highly transient peak pressure events.
    In the flow phantom analysis, \textit{v}WERP was the most accurate method, followed by STE and then PPE\@.
    Each relative pressure estimator was strongly correlated with ground truth relative pressure values, despite the tendency to underestimate peak relative pressure.
    Patient case results demonstrated that each relative pressure estimator could be feasibly integrated into a clinical workflow.
\end{abstract}


%
%

\section{Introduction}\label{sec:introduction}
\IEEEPARstart{R}{elative} pressure differences are clinically important when assessing the severity of a variety of stenotic cardiovascular diseases, including aortic stenosis~\cite{pibarotImprovingAssessmentAortic2012}, aortic coarctation~\cite{kimAorticCoarctation2020}, and mitral valve stenosis~\cite{otto2020ACCAHA2021}.
Meanwhile, relative pressure appears to play a prognostic role in mitral valve regurgitation~\cite{neussElevatedMitralValve2017}, and a mechanistic role in aortic dissection~\cite{baumlerAssessmentAorticDissection2025}.
Until recently, relative pressure differences could only be evaluated by invasive catheterization, which carries inherent risks as a minor surgery~\cite{anjumTransradialVsTransfemoral2017}, or noninvasively by Doppler echocardiography~\cite{harrisQuantitativeDopplerEchocardiography2016}.
While Doppler echocardiography is a useful diagnostic tool in a subset of disorders, it can fail in some patients and only measures one velocity direction.
Therefore, it is limited to simple pressure models, such as the simplified Bernoulli equation.
Very recent work has pushed the bounds of ultrasound, enabling 4D velocity measurements~\cite{ruischUltrasoundbasedVelocityVector2025}.
4D-flow MRI~\cite{markl4DFlowMRI2012} measures three-dimensional (3D), time-resolved velocity fields throughout the cardiac cycle, capturing rich blood flow information.
This flow information enables the estimation of additional hemodynamic variables, such as wall shear stress, kinetic energy, and relative pressure~\cite{vanooijReproducibilityInterobserverVariability2016,binterTurbulentKineticEnergy2017,azarineFourdimensionalFlowMRI2019}.
Relative pressure estimators that utilize 4D-flow MRI have the potential to increase the diagnostic information available by mapping relative pressure across various portions of the cardiovascular system, such as the cerebrovascular system~\cite{valiSemiautomatedAnalysis4D2019}, the heart~\cite{ha4DFlowMRI2019}, and the aorta.
One-dimensional pressure drops derived from catheter measurements or Doppler echocardiography have been in use for decades, but 4D-flow MRI holds the potential for utilizing four-dimensional relative pressure (3D + time) \emph{fields} as a diagnostic tool.
For instance, recent work indicates that pressure drops may be a predictor of growth in dissection---a smaller pressure drop could indicate higher pressurization, leading to subsequent growth and negative outcomes~\cite{marleviFalseLumenPressure2021}.
Furthermore, computational studies point toward the balance of pressures (which requires a fully 4D relative pressure field for computation) in the dissected aorta potentially being an indicator of aortic function~\cite{xuComputedTomographybasedHemodynamic2021}.
While these developments show promise for clinical translation, work is still required to improve technology---from sequences and imaging techniques to analysis.

Herein, we focus on analyzing the performance of relative pressure estimators in type B aortic dissection (TBAD)~\cite{nienaberAorticDissection2016}.
TBAD arises from a partial-thickness entry tear in the descending aorta, creating a second pathway for blood within a false lumen (FL).
The false lumen often reconnects with the true lumen (TL) distally through an exit tear.
TBAD contains a wide variety of complex flow patterns and challenging geometries, making it a valuable setting for the evaluation of relative pressure estimators.
TBAD contains regions with narrow geometries, highly transient pressure events, and low velocities immediately adjacent to high velocities.
Narrow domains are essential when analyzing spatial resolution limits; highly transient events are valuable when analyzing temporal resolution limits.
Low-velocity regions are helpful when analyzing sensitivity to noise.
Finally, high spatial velocity gradients are particularly challenging when estimating hemodynamic metrics~\cite{francoisFourdimensionalFlowsensitiveMagnetic2013a,callaghanUseMultivelocityEncoding2016}, meriting further study.

Relative pressure estimators are highly dependent on the image quality of a given 4D-flow acquisition~\cite{bertoglioRelativePressureEstimation2018}.
Consequently, it is crucial to stress-test these estimators by perturbing and modifying various image properties (e.g., spatial resolution) and determining their impact on relative pressure estimation accuracy.
Previous work has validated and explored the use of various relative pressure estimators~\cite{nolteValidation4DFlow2021,nath4DflowVPNetDeepConvolutional2023,casas4DFlowMRIbased2016,saittaEvaluation4DFlow2019,bertoglioRelativePressureEstimation2018,marleviEstimationCardiovascularRelative2019}, but additional analysis is necessary---the work above has focused on the accuracy of one-dimensional pressure drops, which do not take full advantage of the rich velocity information provided by 4D-flow.
Furthermore, the spatial distribution of errors requires further study.
Herein, we compare CFD-derived relative pressure fields with estimated relative pressure fields in a novel voxel-to-voxel analysis.
The spatial distribution of errors resulting from synthetic noise is also determined.
Compliant TBAD flow phantoms with 12-point arrays of pressure catheter measurements are utilized as ground truth in an \textit{in vitro} study, enhancing our understanding of the spatial variation of errors in relative pressure estimations derived from 4D-flow acquisitions.
In sum, these analyses will add to our understanding of the spatial variations in relative pressure estimator accuracy rather than the accuracy of one-dimensional pressure drops.

In this paper, we analyzed three relative pressure estimators: virtual Work-Energy Relative Pressure (\textit{v}WERP), the Pressure Poisson Estimator (PPE), and the Stokes Estimator (STE), across \textit{in silico}, \textit{in vitro}, and \textit{in vivo} TBAD datasets.
For the \textit{in silico} analysis, we utilized patient-specific CFD simulations to generate synthetic 4D-flow MRI as input.
The resulting estimated relative pressure fields were compared directly to CFD-derived relative pressure fields.
Spatiotemporal and noise sensitivity analyses were performed for both the relative pressure fields and pressure drops.
For the \textit{in vitro} analysis, we utilized 4D-flow acquired across three different compliant TBAD flow phantom variations with varying FL entry and FL exit tears combined with an array of ground truth pressure catheter measurements~\cite{zimmermannHemodynamicEffectsEntry2023}.
Variations in the flow phantoms allowed for a qualitative analysis of the changing relative pressure fields as the TBAD false lumen tears were modified.
Validation of the relative pressure estimators was performed via comparison to catheter pressure measurements.
Finally, pressure drops estimated in eight patient cases were compared across methods for agreement.
Sources of error were identified, and the relative pressure estimators were compared to determine their pros and cons in different use cases.
%
%
\section{Materials and Methods}\label{sec:matmethods}
\subsection{Relative Pressure Estimation Algorithms}
Relative pressure estimation algorithms utilize full-field 4D-flow MRI velocity measurements as input.
These algorithms can be split into \textit{fully spatial} and \textit{plane-to-plane} estimators.
Fully spatial estimators produce a four-dimensional (3D + time) relative pressure field and include PPE and STE\@.
Plane-to-plane estimators produce a time-resolved relative pressure drop between two user-defined planes (e.g., the pressure drop from the aortic root to the FL exit tear throughout one cardiac cycle) and include \textit{v}WERP\@.
The relative pressure estimation algorithms used herein are derived from the incompressible Navier-Stokes equations~\cite{kunduFluidMechanics2025} given by~\eqref{eq:navier_stokes} and~\eqref{eq:continuity}
\begin{align}
    \rho\frac{\partial \vb*{u}}{\partial t} + \rho(\vb*{u}\cdot\nabla)\vb*{u} & = -\nabla p + \mu\Delta\vb*{u} \quad \text{on } \Omega \label{eq:navier_stokes}                           \\
    \nabla\cdot\vb*{u}                                                        & =                                                        0 \quad \text{on } \Omega, \label{eq:continuity}
\end{align}
where \( \Omega \) denotes the spatial domain.
Equation~\eqref{eq:navier_stokes} can be rearranged to estimate pressure gradients by plugging in the measured velocity field, \(\vb*{u}_m\).
\begin{equation}\label{eq:measured_pressure_gradient}
    R_m = -\rho\left(\frac{\partial \vb*{u}_m}{\partial t} + (\vb*{u}_m\cdot\nabla)\vb*{u}_m\right)+\mu\nabla^2\vb*{u}_m \quad \text{on } \Omega
\end{equation}

\subsubsection{Pressure Poisson Estimator (PPE)}
The pressure Poisson estimator~\cite{bertoglioRelativePressureEstimation2018,ebbersNoninvasiveMeasurementTimeVarying2002} is derived by taking the divergence of~\eqref{eq:measured_pressure_gradient} and subsequently solving the resulting Poisson equation with Neumann boundary conditions.
\begin{align}
    \nabla^2\hat{p}           & = \nabla\cdot R_m \quad \text{on } \Omega \label{eq:ppe}              \\
    \nabla\hat{p}\cdot\vb*{n} & = R_m \cdot\vb*{n} \quad \text{on } \partial\Omega, \label{eq:ppe_bc}
\end{align}
where \( \partial\Omega \) is the boundary of the spatial domain and \( \nabla \hat{p} \) is the estimated pressure gradient field.

\subsubsection{Stokes Estimator (STE)}
The Stokes estimator~\cite{svihlovaDeterminationPressureData2016,svihlovaDeterminationPressureData2017} utilizes a divergence-free virtual field, \(\tilde{\vb*{u}}\), as a regularizer~\eqref{eq:stokes_estimator}.
\begin{align}\label{eq:stokes_estimator}
    \nabla\hat{p} - \nabla^2\tilde{\vb*{u}} & = R_m \quad \text{on } \Omega       \\
    \nabla\cdot\tilde{\vb*{u}}              & = 0 \quad \text{on } \Omega         \\
    \tilde{\vb*{u}}                         & = 0 \quad \text{on } \partial\Omega
\end{align}
It does not require higher-order derivatives and has been reported to handle boundaries better because it does not prescribe boundary conditions directly on the pressure field~\cite{nolteValidation4DFlow2021}, as PPE does.
\subsubsection{Virtual Work-Energy Relative Pressure (\textit{v}WERP)}
Virtual work-energy relative pressure is derived from virtual work-energy equations and is based on the principle of conservation of energy.
The virtual work-energy equation is given by multiplying~\eqref{eq:navier_stokes} by a virtual velocity field, \(\vb*{w}\), and integrating over the domain \(\Omega \) between two planes.
\begin{equation}\label{eq:virtual_work_energy}
    \begin{aligned}
        \int_{\Omega}\rho\frac{\partial \vb*{u}_m}{\partial t}\cdot\vb*{w}\,d\Omega + \int_{\Omega}\rho(\vb*{u}_m\cdot\nabla\vb*{u}_m)\cdot\vb*{w}\,d\Omega \\
        - \int_{\Omega}\mu\Delta \vb*{u}_m \cdot \vb*{w}\,d\Omega + \int_{\Omega}\nabla p\cdot\vb*{w}\,d\Omega = 0
    \end{aligned}
\end{equation}
The virtual velocity field \(\vb*{w}\) is constructed by solving a Stokes flow problem over the user-specified flow domain.
After integrating and rearranging~\eqref{eq:virtual_work_energy}, (the details of which can be found in~\cite{marleviEstimationCardiovascularRelative2019}), the pressure drop between two planes is given by
\begin{equation}
    \Delta p = \frac{-1}{Q}\left(\frac{\partial}{\partial t}K_e + A_e + V_e\right)
\end{equation}
where \(K_e\) denotes the kinetic energy held within the blood flow, \(A_e\) denotes the rate of energy entering/leaving the system, and \(V_e\) is the rate of energy dissipation due to viscous forces.
\(Q\) is the volumetric flow rate through the inflow plane of the virtual field\@.
\textit{v}WERP has shown substantial improvement over older plane-to-plane estimators, such as the unsteady Bernoulli equation~\cite{nguyenPressureDifferenceEstimation2019} and the original WERP~\cite{marleviEstimationCardiovascularRelative2019}\@.

\subsection{Implementation of Relative Pressure Estimation Methods}
STE, PPE, and \textit{v}WERP were all implemented using in-house MATLAB code.
All methods utilized a midpoint temporal discretization scheme and second-order spatial derivatives where applicable.
The relevant equations for STE and \textit{v}WERP were discretized with a finite difference method (FDM) that used the measured velocity values directly as input (for details, see the supplement of~\cite{marleviEstimationCardiovascularRelative2019}).
In short, a staggered-grid approach was used.
The velocity and pressure nodes were defined on superimposed grids, with pressure solved at the intersection points.
Both STE and \textit{v}WERP require solving a saddle-point problem.
This was achieved using flexible GMRES right preconditioned with the BFBT block preconditioner (with sub-blocks approximated using algebraic multigrid).
PPE was discretized using a second-order accurate finite volume method (FVM) which was mathematically equivalent to a cell-centered FDM\@.
For PPE, the FVM was chosen over an FDM to eliminate error introduced during boundary normal vector estimation when prescribing Neumann boundary conditions~\eqref{eq:ppe_bc} on the pressure field~\cite{lechnerImagebasedMappingRegional} since all of the boundary normals would then be from the set of Cartesian unit vectors, rather than estimations of the true continuous domain boundary.
Both the FDM (STE) and FVM (PPE) approaches are unique in that they operate directly on the image data---whereas previous implementations~\cite{nolteValidation4DFlow2021} of STE and PPE require a meshing step, our implementations only require a segmentation.
To our knowledge, this is the first time that STE and PPE have been implemented as FDM/FVM.

Finally, the velocity input data were upsampled in all cases.
The primary reason for upsampling was to make all techniques more robust to user-defined masks, which may contain areas primarily filled with border voxels instead of internal voxels.
This can present potential issues, as the grids used for pressure computations are shared with the underlying velocity image.
As numerical solves for all methods are discretized according to the image data, these user masks can yield ill-posed problems, which can be avoided by upsampling.
Specifically, user-defined masks may contain portions of the domain only two voxels thick, meaning finite difference stencils have insufficient nodes to estimate derivatives.
Additionally, in these regions, if Dirichlet conditions are applied, staggered pressure schemes used in \textit{v}WERP and STE may violate discrete inf-sup conditions.
Upsampling enhances the computational domain, filling in additional interior nodes and providing more robustness to user-defined domains.
It should be noted that upsampling makes minimal impact on estimation accuracy in the aorta since the relative pressure solutions are sufficiently smooth; however, upsampling is beneficial to estimator accuracy in other domains, such as the cerebrovasculature~\cite{ferdianCerebrovascularSuperresolution4D2023}.
Importantly, all velocity gradients, Laplacians, and temporal derivatives were calculated using the raw velocity input before upsampling, to avoid potential biases due to interpolation.
For the \textit{in silico} tests, the 3.0, 2.0, and 1.5 mm images were upsampled by factors of 4, 4, and 2, respectively. This led to effective resolutions of 0.75, 0.50, and 0.75 mm.

\subsection{In Silico Analysis}
\subsubsection{Model Generation}\label{sec:cfd_generation}
A patient-specific CFD model of a TBAD patient was used as a test bed to analyze PPE, STE, and \textit{v}WERP\@.
The CFD model enabled the generation of synthetic 4D-flow images at varying spatiotemporal resolutions and noise levels.
Model geometry was sourced from a clinical CT, segmented in 3D Slicer~\cite{joleszIntraoperativeImagingImageGuided2014}, and meshed with 1.1 mm elements and a 0.9 mm boundary layer in Simmetrix SimModeler~\cite{SimModeler}.
The resulting mesh had a total of 2,564,776 linear elements.
The boundary layer was three elements thick and followed a geometric growth pattern: the outermost layer was 0.15 mm thick, the next layer was 0.25 mm thick, and the innermost layer was 0.5 mm thick.
The simulated time resolution was 1.0 ms and there were 1,852,032 degrees of freedom.
Boundary conditions were prescribed using a 4D-flow MRI scan (3T;\ voxel size = \({\sim}1.5 \text{ mm} \times 1.5 \text{ mm} \times 2.5 \text{ mm}\); temporal resolution = 47 ms, VENC = 2.0 m/s) of the same patient.
Inflow and outflow boundary conditions were imposed at the aortic root, brachiocephalic artery, left common carotid artery, left subclavian artery, and true lumen outlet based on flow rates interpolated from 4D-flow to 1.0 ms resolution.
Flow resistance was estimated by measuring the area of each outlet.
The aortic wall was assumed to be completely rigid.
A single cardiac cycle was simulated in \(\mathcal{C}\)Heart~\cite{leeMultiphysicsComputationalModeling2016} using a stabilized Navier-Stokes finite element scheme (see~\cite{hessenthalerValidationNonconformingMonolithic2017} for details).
Use of stabilized methods is common in cardiovascular CFD, but does allow for violations of weak mass conservation (through so-called pressure stabilization), especially on coarse grids.
In theory, this could impact velocity-based pressure estimations relying on incompressibility assumptions---though this has not been observed in our studies.
The solved pressure field, \(p_{\text{CFD}}\), was interpolated from the unstructured CFD mesh onto isotropic 0.50 mm and 0.75 mm grids at a 20 ms sampling rate.
These relative pressure interpolations served as the ground truth.
\begin{figure}[h]
    \centering
    \includegraphics{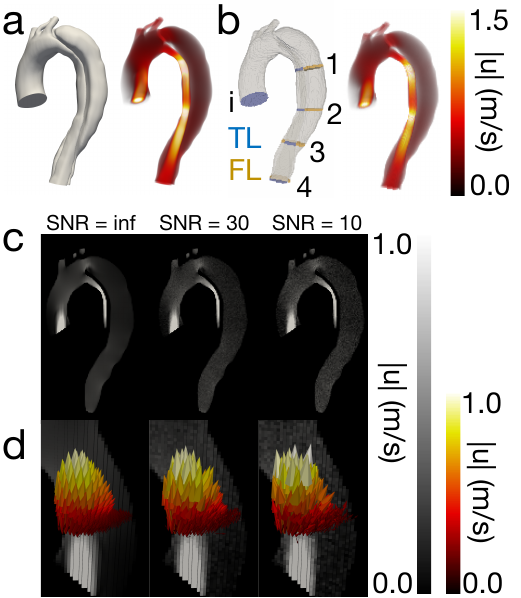}
    \caption{Overview of \textit{in silico} model at peak systolic flow\@. (a) Solved velocity field on unstructured FEM mesh\@. (b) Solved velocity field after interpolation to a voxelized domain. Planes used for pressure drop analysis are shown in blue and gold\@. (c) Noisy velocity magnitude images at tested SNR levels. Non-fluid regions are set to zero to aid comparison. (d) Noisy velocity vector planes in the ascending aorta at tested SNR levels.}\label{fig:cfd_overview}
\end{figure}

\subsubsection{Spatiotemporal Analysis}
The solved velocity field, \(\vb*{u}_{\text{CFD}}\), was interpolated from the unstructured CFD mesh onto a variety of regular, isotropic grids to mimic MRI image format (\autoref{fig:cfd_overview}ab).
Interpolations were made onto 1.5, 2.0, and 3.0 mm grids at 20 ms, 40 ms, and 60 ms sampling rates for a total of nine spatiotemporal combinations.
This allowed for a spatiotemporal analysis of PPE, STE, and \textit{v}WERP to evaluate the robustness of each relative pressure estimation method across a variety of resolutions.
Performance was analyzed by propagating each of the \(\vb*{u}_{\text{CFD}}\) fields through each estimator and comparing the estimated pressure results (\(p_{\text{PPE}}\), \(p_{\text{STE}}\)) with the interpolated ground truth pressure solution, \(p_{\text{CFD}}\), voxel by voxel.
Since \textit{v}WERP is a plane-to-plane estimator, \textit{v}WERP-derived pressure drops were compared with pressure drops sourced from the interpolated \(p_{\text{CFD}}\) rather than directly with the full relative pressure field.
Nine unique planes (for eight separate pressure drops) were generated for this purpose (\autoref{fig:cfd_overview}b).

PPE and STE produce spatially comprehensive relative pressure estimations (\(p_\text{STE}\), \(p_\text{PPE}\)) that were compared directly with \(p_{\text{CFD}}\) images.
Relative error over the entire spatial domain was evaluated at each timestep with~\eqref{eq:rel_error_time} and across the entire spatiotemporal domain with~\eqref{eq:rel_error_all}, producing a single scalar value per dataset.\footnote{Here \( \| \cdot \|_{0,p} \) refers to the natural norm for the \(L^p\) Banach space over the appropriate flow domain.}
\begin{equation} \label{eq:rel_error_time}
    \epsilon_{\text{rel}}(t) = \frac{\left\lVert p_{\text{CFD}}(t) - \hat{p}(t)\right\rVert_{0,2}}{\left\lVert p_{\text{CFD}} \right\rVert_{0,\infty}}
\end{equation}
\begin{equation} \label{eq:rel_error_all}
    \varepsilon_{\text{rel}} = \frac{\left\lVert p_{\text{CFD}} - \hat{p}\right\rVert_{0,2}}{\left\lVert p_{\text{CFD}} \right\rVert_{0,2}}
\end{equation}
In addition to the relative error metrics, correlation and Bland--Altman plots were generated.

\textit{v}WERP pressure drop estimates were compared directly with pressure drops selected from \(p_{\text{CFD}}\).
Planes were drawn in ParaView~\cite{ahrensParaViewEndUserTool2005}.
Pressure drops selected from \(p_{\text{STE}}\) and \(p_{\text{PPE}}\) were compared with pressure drops selected from \(p_{\text{CFD}}\) as well.
As in the fully spatial evaluation, linear regression plots were generated.

\subsubsection{Noise Sensitivity Analysis}
To assess the impact of noise on relative pressure estimate accuracy, random noise fields~\cite{ferdian4DFlowNetSuperResolution4D2020} were added to interpolated CFD velocity fields before evaluating the performance of PPE, STE, and \textit{v}WERP\@.
Noise was added as follows: velocity fields were first converted to phase data by specifying a synthetic VENC that was 1.1 times the maximum velocity component across the entire time-resolved velocity field.
Synthetic magnitude images were created by multiplying the voxelized CFD mask by a non-zero constant (leaving the non-fluid region equal to 0).
The synthetic magnitude and phase images were combined to create complex images that were subsequently transformed to \textit{k}-space via the Fast Fourier Transform.
Zero-mean, complex Gaussian noise was added to \textit{k}-space.
The variance of the complex Gaussian was defined to achieve a specified SNR using the equation: \(\text{SNR}_{\text{dB}} = 10\log{\frac{P_x}{P_n}}\), where \(P_x\) is the power of the signal and \(P_n\) is the power of the noise (which is equivalent to the variance for white Gaussian noise).
Finally, the \textit{k}-space data were transformed back to image space via the inverse Fourier Transform, and the phase images were converted back to velocity images using the specified VENC\@.
Twenty-five realizations of SNR = 10 noise fields and SNR = 30 noise fields (\autoref{fig:cfd_overview}c) were calculated for a total of 50 realizations per resolution, resulting in 450 realizations in total.
The SNR = inf case was performed implicitly during the spatiotemporal analysis.
The same error metrics used in the previous spatiotemporal analysis were used for the noise sensitivity analysis, with some modifications to the correlation analysis---instead of performing a linear regression over one relative pressure field, linear regressions were performed over all 25 noise realizations at once.
Error averages, minima, and maxima were calculated for each spatiotemporal and noise combination (e.g., the average error was calculated over the 25 realizations of the 1.5 mm \(\times \) 40 ms and SNR = 30 combination).
Finally, pressure drops were evaluated for their robustness to noise with~\eqref{eq:mean_deviation} and~\eqref{eq:max_deviation}.
Equations~\eqref{eq:mean_deviation} and~\eqref{eq:max_deviation} are referred to in the text as ``deviations'' because they are designed to measure the spread of the noisy pressure estimations from the baseline (no noise) estimations.
\begin{equation} \label{eq:mean_deviation}
    e_{\Sigma} = \frac{\left\lVert\Delta p_{\text{original}}(t)-\Delta p_{\text{corrupted}}(t)\right\rVert_{0,1}}{\left\lVert\Delta p_{\text{original}}(t)\right\rVert_{0,1}}
\end{equation}
\begin{equation} \label{eq:max_deviation}
    e_{\text{max}} = \frac{\left\lVert\Delta p_{\text{original}}(t_n)-\Delta p_{\text{corrupted}}(t_n)\right\rVert_{0,\infty}}{\left\lVert\Delta p_{\text{original}}(t_n)\right\rVert_{0,\infty}}
\end{equation}

\subsection{Flow Phantoms}
PPE, STE, and \textit{v}WERP were validated with an \textit{in vitro} flow setup~\cite{zimmermannHemodynamicEffectsEntry2023}.
4D-flow MRI was acquired, and twelve pressure catheter measurements per TBAD tear variation served as ground truth.
\subsubsection{Flow Setup}
\begin{figure}[h]
    \centering
    \includegraphics{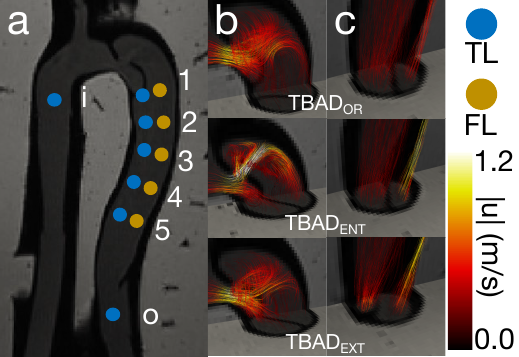}
    \caption{Overview of \textit{in vitro} flow setup\@. (a) Catheter measurement points and cut planes across TBAD tear variations\@. (b) Entry tears across all TBAD tear variations\@. (c) Exit tears across all TBAD tear variations.}\label{fig:judith_phantom}
\end{figure}

A full description of the flow phantom can be found in~\cite{zimmermannHemodynamicEffectsEntry2023} (pump model, exact flow specifications, etc.).
In brief, the flow phantom geometry was designed as an embedded compliant 3D-printed TBAD model derived from a clinical CT exam.
Physiologically accurate pulsatile flow and pressure boundary conditions were applied. 4D-flow MRI (3T;\ voxel size = \(1.5\times1.5\times1.5\) mm; temporal resolution = 50 ms; VENC = 1.2--1.7 m/s; parallel imaging (GRAPPA, R = 2)) was acquired across three phantom variations: one baseline phantom (TBAD\textsubscript{OR}) and two additional variations with identical anatomy but with either a reduced-size FL entry tear (TBAD\textsubscript{ENT}) or a reduced-size FL exit tear (TBAD\textsubscript{EXT}) shown in \autoref{fig:judith_phantom}bc.
Twelve catheter pressure measurements were acquired at consistent locations across each TBAD tear variation, for a total of 36 unique measurements (\autoref{fig:judith_phantom}a).

\subsubsection{Validation Methodology}
Eleven time-resolved pressure drops derived from the twelve pressure catheter locations served as the ground truth.
To validate PPE and STE, the entire relative pressure field was calculated (\(p_{\text{STE}}, p_{\text{PPE}}\)), and individual pressure drops were found by spatially averaging the relative pressure values at each catheterization plane (i.e., the specified inlet and outlet planes) and calculating the difference at each time point to produce temporally resolved, one-dimensional pressure drops\@.
\textit{v}WERP, being a plane-to-plane estimator, natively calculated these one-dimensional pressure drops, so no additional post-processing was required.
Correlation and Bland--Altman plots were created to analyze the performance of each relative pressure estimator.

\subsubsection{Sensitivity Analysis}
As with the CFD model, noise was added to the velocity fields prior to pressure estimation to evaluate each method's robustness to noise.
Noise, while minimal, was already present in the acquired 4D-flow, so any additional noise is denoted as \(\Delta\text{SNR}\).
Complex Gaussian noise was added in \textit{k}-space following the same procedure as in the \textit{in silico} analysis.
The only difference was that the original acquired magnitude images were used to create \textit{k}-space, rather than a synthetic mask-based magnitude image.
This difference made the characteristics of the noise more representative of clinical conditions---SNR was subtly varied throughout the image when compared to using a magnitude image with a constant value throughout the domain.
Each TBAD tear variation was evaluated over 50 realizations with \(\Delta \)SNR = 30 and \(\Delta \)SNR = 10, for a total of 300 noise realizations per relative pressure estimation method.
Pressure drop deviations were calculated for each unique noisy velocity field realization.
This included measures of mean deviation~\eqref{eq:mean_deviation} and max deviation~\eqref{eq:max_deviation}.
Error averages, minimums, and maximums were recorded across each of the 50 realization combinations (e.g., \(\text{TBAD}_{\text{ENT}}\) at \(\Delta \)SNR = 10).

\subsection{Patient Cases}
\begin{figure}[h]
    \centering
    \includegraphics{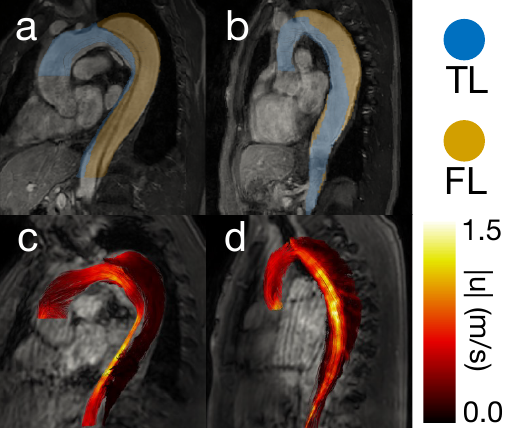}
    \caption{Two representative patient cases.\ (a) High-growth patient case MRA with 3D segmentation overlaid. The TL segmentation is blue, and the FL segmentation is gold.\ (b) Low-growth patient case MRA with 3D segmentation overlaid.\ (c) High-growth patient case 4D-flow magnitude with velocity streamlines at peak systole overlaid.\ (d) Low-growth patient case 4D-flow magnitude with velocity streamlines at peak systole overlaid.}\label{fig:patient_overview}
\end{figure}

The final analysis utilized eight patient cases (\autoref{fig:patient_overview}ab) to assess the clinical feasibility of relative pressure estimation methods and to qualitatively evaluate relative pressure trends against what is known in the literature, including our knowledge of FL growth rate (FL GR)~\cite{marleviFalseLumenPressure2021}.
4D-flow was acquired on 3T MRI scanners (MR750, GE Medical Systems, Milwaukee, WI, USA;\ Ingenia, Philips, Best, Netherlands) with the following scan parameters: voxel size = \({\sim}1.5 \text{ mm} \times 1.5 \text{ mm} \times 2.5 \text{ mm}\), average temporal resolution = 47 ms, VENC = 2.0 m/s, parallel imaging with SENSE (R = 2 $\times$ 2).
Agreement between the relative pressure estimators was evaluated.
4D-flow was acquired from a previous IRB-approved study~\cite{burrisFalseLumenEjection2020} (HUM00120679).
This patient cohort was also used in Marlevi \textit{et al.}~\cite{marleviFalseLumenPressure2021} to analyze \textit{v}WERP; here we reuse this cohort to compare \textit{v}WERP to STE and PPE.
The data were segmented by thresholding patient MRA data and subsequently interpolating the resulting segmentation onto the 4D-flow image space (\autoref{fig:patient_overview}cd).
Inlet and outlet planes were generated in a 4D-flow MATLAB toolbox~\cite{soteloJulioSoteloParraguez4DFlowMatlabToolbox2024}.
FL GR was measured clinically as the annualized change in maximum aortic diameter (mm/yr) in the descending aorta.
Each relative pressure estimator was run on all eight \textit{in vivo} cases.
The relative pressure fields calculated by the fully spatial relative pressure estimators were analyzed for their coherence.
The 4D relative pressure fields were also converted to pressure drops (inlet--TL outlet and inlet--FL outlet) to allow for comparisons with \textit{v}WERP\@.

%
%
\section{Results}\label{sec:results}
\subsection{Evaluation of \textit{In Silico} TBAD Model}
\subsubsection{Spatiotemporal Analysis}
\begin{figure}[h]
    \centering
    \includegraphics{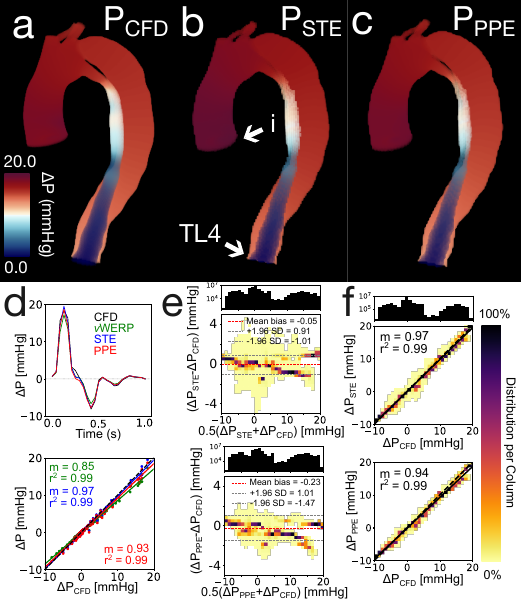}
    \caption{Overview of the \textit{in silico} 1.5 mm \(\times \) 40 ms results.\ (a) CFD ground truth relative pressure field at peak systole\@. (b) STE relative pressure field at peak systole. Aortic inlet (i) and true lumen outlet (TL4) are marked with arrows for subfigure d\@. (c) PPE relative pressure field at peak systole. (d) Pressure drop from aortic inlet to TL outlet and linear regression for each relative pressure estimation method\@. (e) Bland--Altman plots for STE and PPE\@. The heatmaps indicate the distribution of values in each 2D bin. The number of points in each 2D bin is normalized to the number of values in the entire column. (f) Linear regressions performed on STE and PPE relative pressure fields across all timesteps. The heatmaps indicate the distribution of relative pressure values in each 2D bin. The number of points in each 2D bin is normalized to the number of points in the entire column. The 1D histogram above the heatmap correlation plots illustrates the number of points present within each column.}\label{fig:in_silico_baseline}
\end{figure}

The spatiotemporal analysis was performed on nine unique spatiotemporal resolutions.
The 1.5 mm \(\times \) 40 ms results are visualized in \autoref{fig:in_silico_baseline} as a representative example.
\autoref{fig:in_silico_baseline}abc presents the CFD, STE, and PPE relative pressure fields, respectively, at peak systole.
Relative error values are recorded in~\autoref{tab:relative_error_no_noise} and the Bland--Altman results are recorded in~\autoref{tab:in_silico_blandaltman_no_noise}.
\autoref{fig:in_silico_baseline}e shows the Bland--Altman plots for STE and PPE versus the CFD ground truth.
\autoref{fig:in_silico_baseline}f shows the correlation between the relative pressure fields calculated by STE and PPE versus the CFD ground truth.
Throughout the text, the binned Bland--Altman plots and linear regressions are normalized as follows: the number of points in each 2D bin is divided by the number of points in the entire column, and the small 1D histogram at the top of each binned regression indicates what this per-column normalization factor was.
This was done because the vast majority of the values cluster around 0 mmHg.
The estimated relative pressure fields were highly consistent---both \(p_{\text{STE}}\) and \(p_{\text{PPE}}\) were very strongly correlated with \(p_{\text{CFD}}\).
PPE consistently slightly underestimated the ground truth relative pressure compared to STE, which overestimated ground truth pressures\@.

\autoref{fig:in_silico_baseline}d shows the pressure drop from the aortic inlet to the TL outlet as calculated by \textit{v}WERP, STE, and PPE\@.
\autoref{tab:in_silico_dp} summarizes the pressure drop linear regression results for the spatiotemporal analysis.
All correlation coefficients were near unity for the 1.5 mm and 2.0 mm images, but degraded substantially for the 3.0 mm images.
PPE correlation coefficients decreased the most drastically for the 3.0 mm cases.

\begin{table}[h]
    \centering
    \caption{In Silico Total Relative Error (\(\varepsilon_{\text{rel}}\)) for Fully Spatial Relative Pressure Fields}
    \begin{tabularx}{\linewidth}{ @{\extracolsep{\fill}} *{7}c @{} }\toprule
               & \multicolumn{3}{c}{STE} & \multicolumn{3}{c}{PPE}                                            \\
        \midrule \midrule
               & 20 ms                   & 40 ms                   & 60 ms   & 20 ms    & 40 ms    & 60 ms    \\
        \cmidrule{2-4} \cmidrule{5-7}
        1.5 mm & 3.33\%                  & 8.76\%                  & 16.89\% & 8.30\%   & 11.98\%  & 16.69\%  \\
        2.0 mm & 19.57\%                 & 18.34\%                 & 24.83\% & 6.79\%   & 8.51\%   & 13.69\%  \\
        3.0 mm & 78.98\%                 & 80.43\%                 & 85.87\% & 101.76\% & 103.10\% & 110.15\% \\
        \bottomrule
    \end{tabularx}\label{tab:relative_error_no_noise}
\end{table}

\begin{table}[h]
    \centering
    \caption{Mean Bias and Standard Deviations for In Silico Fully Spatial Relative Pressure Fields}
    \begin{tabularx}{\linewidth}{ @{\extracolsep{\fill}} *{4}c @{} }\toprule
        STE    & 20 ms               & 40 ms              & 60 ms               \\
        \cmidrule{2-4}
        1.5 mm & \(0.03 \pm 0.19\)   & \(-0.05 \pm 0.49\) & \(0.14 \pm 0.80\)   \\
        2.0 mm & \(0.56 \pm 0.96\)   & \(0.48 \pm 0.91\)  & \(0.64 \pm 1.01\)   \\
        3.0 mm & \(-2.15 \pm 3.93\)  & \(-2.21 \pm 3.93\) & \(-1.88 \pm 3.69\)  \\
        \midrule\midrule
        PPE    & 20 ms               & 40 ms              & 60 ms               \\
        \cmidrule{2-4}
        1.5 mm & \(-0.16 \pm 0.45 \) & \(-0.23 \pm 0.63\) & \(-0.02 \pm 0.81\)  \\
        2.0 mm & \(0.12 \pm 0.37 \)  & \(0.04 \pm 0.48\)  & \(0.24 \pm 0.62\)   \\
        3.0 mm & \(-2.80 \pm 5.06 \) & \(-2.86 \pm 5.04\) & \(-2.49 \pm 4.71 \) \\
        \bottomrule
    \end{tabularx}\label{tab:in_silico_blandaltman_no_noise}
\end{table}

\begin{table}[t]
    \centering
    \caption{In Silico Pressure Drop Slopes and Correlations}
    \begin{tabularx}{\linewidth}{ @{\extracolsep{\fill}} *{11}c @{} }\toprule
                                &         & \multicolumn{3}{c}{\textit{v}WERP} & \multicolumn{3}{c}{STE} & \multicolumn{3}{c}{PPE}                                                 \\
        \midrule \midrule
                                &         & 20 ms                              & 40 ms                   & 60 ms                   & 20 ms & 40 ms & 60 ms & 20 ms & 40 ms & 60 ms \\
        \cmidrule{3-5}\cmidrule{6-8}\cmidrule{9-11}
        \multirow{2}{*}{1.5 mm} & \(m\)   & 0.86                               & 0.85                    & 0.92                    & 0.99  & 0.92  & 1.04  & 0.94  & 0.93  & 0.99  \\
                                & \(r^2\) & 1.00                               & 0.99                    & 0.98                    & 1.00  & 0.99  & 0.98  & 1.00  & 0.99  & 0.97  \\
        \multirow{2}{*}{2.0 mm} & \(m\)   & 0.89                               & 0.88                    & 0.96                    & 1.09  & 1.08  & 1.17  & 1.01  & 0.99  & 1.06  \\
                                & \(r^2\) & 0.98                               & 0.98                    & 0.99                    & 0.98  & 0.98  & 0.99  & 0.99  & 0.99  & 0.99  \\
        \multirow{2}{*}{3.0 mm} & \(m\)   & 0.36                               & 0.35                    & 0.36                    & 0.54  & 0.52  & 0.54  & 0.42  & 0.41  & 0.40  \\
                                & \(r^2\) & 0.36                               & 0.34                    & 0.30                    & 0.39  & 0.37  & 0.33  & 0.21  & 0.19  & 0.16  \\
        \bottomrule
    \end{tabularx}\label{tab:in_silico_dp}
\end{table}

\subsubsection{Sensitivity Analysis Results}
\begin{figure}[h]
    \centering
    \includegraphics{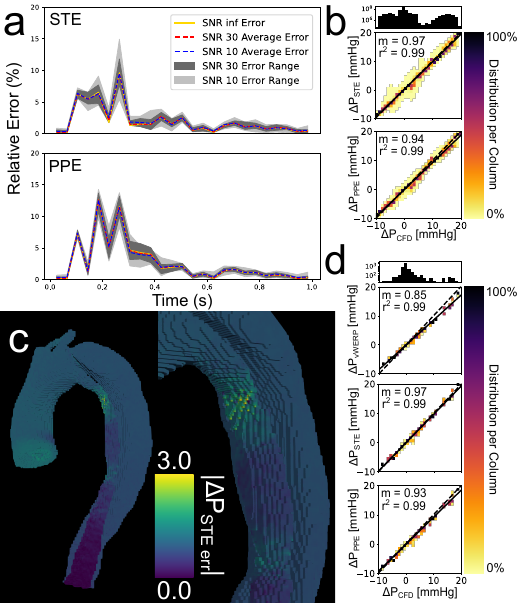}
    \caption{Overview of \textit{in silico} 1.5 mm \(\times \) 40 ms sensitivity analysis\@. (a) Relative error over cardiac cycle\@. (b) Correlation heatmaps for the fully spatial analysis over all 25 noise realizations\@. (c) Relative pressure error map for STE at SNR = 10. (d) Correlation heatmaps for the plane-to-plane pressure drop analysis across all 25 noise realizations. The heatmaps indicate the distribution of relative pressure values in each 2D bin. The number of points in each 2D bin is normalized to the number of points in the entire column. The 1D histogram above the heatmap correlation plots illustrates the number of points present within each column.}\label{fig:in_silico_sensitivity}
\end{figure}

The sensitivity analysis for the 1.5 mm \(\times \) 40 ms resolution is summarized in \autoref{fig:in_silico_sensitivity}.
The relative error over the cardiac cycle~\eqref{eq:rel_error_time} is shown in \autoref{fig:in_silico_sensitivity}a for STE and PPE\@.
There was not a consistent trend in the relative errors of STE and PPE---STE outperformed PPE across the 1.5 mm cases, but underperformed in the 2.0 and 3.0 mm cases in terms of noise resistance.
Correlation heatmaps for the fully spatial analysis and the plane-to-plane pressure drop analysis at SNR = 10 are shown in \autoref{fig:in_silico_sensitivity}b and d, respectively.
A relative pressure error map for an SNR = 10 1.5 mm \(\times \) 40 ms case is shown in \autoref{fig:in_silico_sensitivity}c.
Errors were lowest in the distal true lumen.
In general, errors grew when moving proximally from the true lumen outlet.
The exception to this trend was in the extremely narrow region of the proximal true lumen, which is shown in more detail in \autoref{fig:in_silico_sensitivity}c (right).
The range of total relative errors~\eqref{eq:rel_error_all} is shown in \autoref{tab:relative_error_sensitivity}.
Averaged total relative errors stayed roughly the same as noise was increased; the maximum change in average total relative error was 1.30\% for PPE 2.0 mm \(\times \) 60 ms when increasing from SNR = inf to SNR = 10.
Relative errors for STE were lower than for PPE for the 1.5 mm and 3.0 mm resolutions;\ however, STE generally had larger error ranges, indicating that it was more sensitive to noise.

\begin{table}[h]
    \centering
    \caption{In Silico Total Relative Error (\(\varepsilon_{\text{rel}}\)) Range for Sensitivity Analysis}
    \begin{tabular*}{\linewidth}{ @{\extracolsep{\fill}} c *{7}c @{} }\toprule
        & & \multicolumn{3}{c}{STE} & \multicolumn{3}{c}{PPE}  \\
        \midrule \midrule
        SNR & dx/dt & 20 ms & 40 ms & 60 ms  & 20 ms & 40 ms & 60 ms   \\
        \cmidrule{3-5} \cmidrule{6-8}
        \multirow{3}{*}{30}  & 1.5 mm & 2.08\% & 1.74\% & 1.48\% & 1.61\% & 2.08\% & 2.34\%  \\
        & 2.0 mm & 3.43\% & 4.48\% & 5.20\% & 2.72\% & 2.52\% & 3.14\% \\
        & 3.0 mm & 8.04\% & 12.50\% & 16.18\% & 11.75\% & 12.97\% & 16.90\% \\
        \midrule
        \multirow{3}{*}{10}  & 1.5 mm & 1.28\% & 3.24\% & 3.37\% & 2.76\% & 4.00\% & 4.20\% \\
        & 2.0 mm & 7.06\% & 5.14\% & 5.69\% & 2.59\% & 2.78\% & 6.15\% \\
        & 3.0 mm & 23.97\% & 28.18\% & 30.06\% & 15.42\% & 16.99\% & 28.98\% \\
        \bottomrule
    \end{tabular*}\label{tab:relative_error_sensitivity}
\end{table}

In general, all methods were resilient to noise; only the 3.0 mm relative pressure traces had a relative error range larger than 10\% due to noise introduced into \(\vb*{u}_\text{CFD}\).
Furthermore, the calculated deviation values indicated that the spread from the baseline (no noise) cases increased substantially for the 3.0 mm cases.
Across all methods, the deviation values for the 1.5 mm and 2.0 mm cases remained below 4\% in almost all cases and frequently below 2\%.
Of the 3.0 mm cases, the \textit{v}WERP estimations showed the highest deviations from the baseline case, followed by STE, and then PPE.

\subsection{Flow Phantom Analysis}

\begin{figure}[h]
    \centering
    \includegraphics{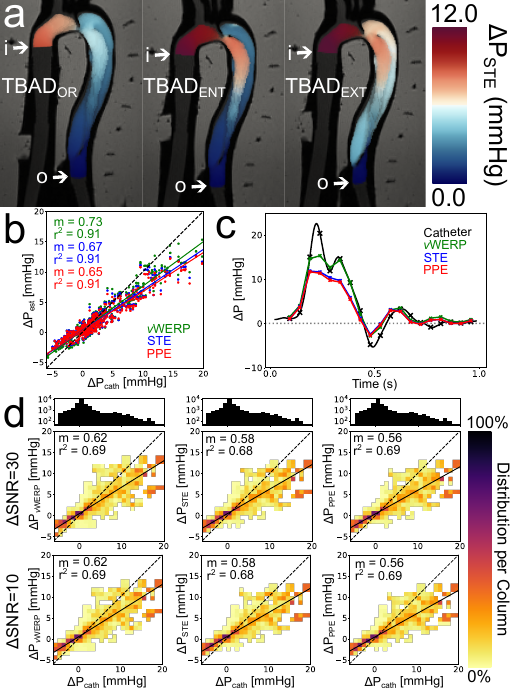}
    \caption{Overview of \textit{in vitro} flow phantom analysis\@. (a) \textit{In vitro} model STE relative pressure fields at peak systole. The annotations i and o indicate the inlet and outlet, respectively, that bounded the velocity domain when solving for relative pressure. PPE relative pressure fields are not shown, but were nearly identical\@. (b) Linear regression across all models and planes with no added noise\@. (c) Example relative pressure trace over the cardiac cycle. The pressure drop was calculated from the model inlet to the model outlet (i to o in subfigure 6a)\@. The ``x'' markers indicate where 4D-flow timeframes were present\@. (d) Correlation heatmaps for each method. Linear regressions were run over all 50 noise realizations per TBAD tear variation for a total of 150 unique datasets per correlation heatmap. The heatmaps indicate the distribution of relative pressure values in each 2D bin. The number of points in each 2D bin is normalized to the number of points in the entire column. The 1D histogram above the heatmap correlation plots illustrates the number of points present within each column.}\label{fig:invitro_overview}
\end{figure}

\begin{figure}[h]
    \centering
    \includegraphics{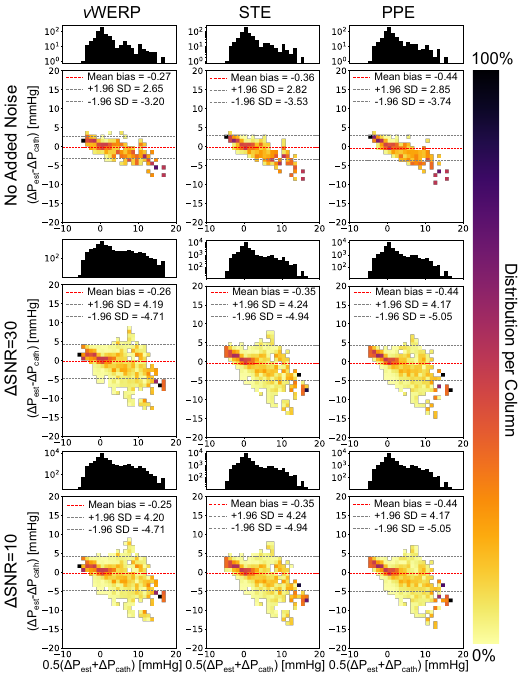}
    \caption{Bland--Altman analysis of \textit{in vitro} relative pressure estimators. The plots shown are composites of all TBAD tear variations and all noise levels, with the bias and LoA lines calculated across all noise realizations at once.}\label{fig:invitro_blandaltman}
\end{figure}

\subsubsection{Validation of Relative Pressure Estimators}
A broad overview of the flow phantom analysis results is given in \autoref{fig:invitro_overview} and \autoref{fig:invitro_blandaltman}.
Qualitatively, the relative pressure fields estimated by STE and PPE demonstrated the expected features relative to each model variation's geometries and resulting velocity fields (\autoref{fig:judith_phantom}).
For instance, the high relative pressure impingement zone immediately following the TBAD\textsubscript{ENT} entry tear in \autoref{fig:invitro_overview}a closely resembles the impingement zone in the corresponding FSI model of TBAD\textsubscript{ENT}, shown in Fig. 4c of Zimmermann \textit{et al.}~\cite{zimmermannHemodynamicEffectsEntry2023}.
Both the STE and PPE relative pressure fields exhibited a large pressure drop across the FL entry tear and higher pressure drops from the aortic inlet (i) to each TL measurement point (TL 1--5) compared to each FL measurement point (FL 1--5)\@.
The regression analysis in \autoref{fig:invitro_overview}b indicated that \textit{v}WERP was the most accurate relative pressure estimator when no noise was added.
A representative pressure drop from the model inlet to the model outlet is illustrated in \autoref{fig:invitro_overview}c to demonstrate this trend.
Across all TBAD tear variations, \textit{v}WERP, STE, and PPE were strongly correlated (\(r^2 = 0.91\)) with catheter measurements, with a tendency to underestimate peak relative pressures.
This trend held across all relative pressure traces.

\subsubsection{Sensitivity Analysis Results}
\autoref{fig:invitro_overview}d illustrates the effect of the \(\Delta \)SNR = 30 and \(\Delta \)SNR = 10 noise levels across all 50 realizations and all models for each relative pressure estimation method.
Noise added to the velocity fields decreased the linear regression slopes by roughly 0.10 for each relative pressure estimator (e.g., from 0.73 at SNR = inf to 0.62 at \(\Delta\)SNR = 30 and \(\Delta\)SNR = 10 for \textit{v}WERP), while the correlations were moderately weaker (\(r^2\) = 0.68--0.69).
The pressure drop mean deviation~\eqref{eq:mean_deviation} and max deviation~\eqref{eq:max_deviation} results for all models are summarized in \autoref{tab:tbad_ent_deviations}.
For the FL and TL rows, deviations were averaged across all numbered pressure drops for that lumen (e.g., aortic inlet to TL1, aortic inlet to TL2, etc.).
As expected, the average deviation increased with noise.
Average deviations were also reliably higher in the true lumen across all models and relative pressure estimation methods, particularly for STE and PPE.
On average, STE had the lowest deviation values, closely followed by \textit{v}WERP. PPE had the highest deviation values.
\begin{table*}[t]
    \centering
    \caption{In Vitro Average Deviations Across All Models}
    \begin{tabularx}{\linewidth}{ @{\extracolsep{\fill}} *{13}c @{} }\toprule
                               & \multicolumn{4}{c}{\textit{v}WERP}   & \multicolumn{4}{c}{STE}              & \multicolumn{4}{c}{PPE}                                                                                                                                                                                                                                   \\
        \midrule \midrule
        TBAD\textsubscript{OR} & \multicolumn{2}{c}{$\Delta$SNR = 30} & \multicolumn{2}{c}{$\Delta$SNR = 10} & \multicolumn{2}{c}{$\Delta$SNR = 30} & \multicolumn{2}{c}{$\Delta$SNR = 10} & \multicolumn{2}{c}{$\Delta$SNR = 30} & \multicolumn{2}{c}{$\Delta$SNR = 10}                                                                                                 \\
        \cmidrule{2-5}\cmidrule{6-9}\cmidrule{10-13}
                               & \(e_{\Sigma}\)                       & \(e_{\max}\)                         & \(e_{\Sigma}\)                       & \(e_{\max}\)                         & \(e_{\Sigma}\)                       & \(e_{\max}\)                         & \(e_{\Sigma}\) & \(e_{\max}\) & \(e_{\Sigma}\) & \(e_{\max}\) & \(e_{\Sigma}\) & \(e_{\max}\) \\
        \cmidrule{2-3}\cmidrule{4-5}\cmidrule{6-7}\cmidrule{8-9}\cmidrule{10-11}\cmidrule{12-13}
        FL                     & 0.58\%                               & 0.49\%                               & 1.04\%                               & 0.85\%                               & 0.46\%                               & 0.35\%                               & 0.74\%         & 0.61\%       & 0.46\%         & 0.38\%       & 0.81\%         & 0.65\%       \\
        TL                     & 1.06\%                               & 0.80\%                               & 1.84\%                               & 1.41\%                               & 0.97\%                               & 0.71\%                               & 1.65\%         & 1.27\%       & 0.97\%         & 0.72\%       & 1.71\%         & 1.35\%       \\
        Outlet                 & 0.53\%                               & 0.38\%                               & 0.92\%                               & 0.67\%                               & 0.43\%                               & 0.30\%                               & 0.73\%         & 0.54\%       & 0.44\%         & 0.35\%       & 0.77\%         & 0.56\%       \\
    \end{tabularx}\label{tab:tbad_or_deviations}
    \centering
    \begin{tabularx}{\linewidth}{ @{\extracolsep{\fill}} *{13}c @{} }
        \midrule \midrule
        TBAD\textsubscript{ENT} & \multicolumn{2}{c}{$\Delta$SNR = 30} & \multicolumn{2}{c}{$\Delta$SNR = 10} & \multicolumn{2}{c}{$\Delta$SNR = 30} & \multicolumn{2}{c}{$\Delta$SNR = 10} & \multicolumn{2}{c}{$\Delta$SNR = 30} & \multicolumn{2}{c}{$\Delta$SNR = 10}                                                                                                 \\
        \cmidrule{2-5}\cmidrule{6-9}\cmidrule{10-13}
                                & \(e_{\Sigma}\)                       & \(e_{\max}\)                         & \(e_{\Sigma}\)                       & \(e_{\max}\)                         & \(e_{\Sigma}\)                       & \(e_{\max}\)                         & \(e_{\Sigma}\) & \(e_{\max}\) & \(e_{\Sigma}\) & \(e_{\max}\) & \(e_{\Sigma}\) & \(e_{\max}\) \\
        \cmidrule{2-3}\cmidrule{4-5}\cmidrule{6-7}\cmidrule{8-9}\cmidrule{10-11}\cmidrule{12-13}
        FL                      & 0.92\%                               & 1.05\%                               & 1.34\%                               & 1.49\%                               & 0.59\%                               & 0.67\%                               & 0.98\%         & 1.13\%       & 1.03\%         & 2.25\%       & 1.50\%         & 2.81\%       \\
        TL                      & 1.34\%                               & 1.14\%                               & 2.33\%                               & 1.90\%                               & 2.47\%                               & 5.26\%                               & 3.49\%         & 6.39\%       & 3.81\%         & 9.63\%       & 5.25\%         & 11.67\%      \\
        Outlet                  & 0.90\%                               & 1.02\%                               & 1.31\%                               & 1.40\%                               & 0.62\%                               & 0.70\%                               & 1.08\%         & 1.20\%       & 1.55\%         & 4.09\%       & 2.13\%         & 4.86\%       \\
    \end{tabularx}\label{tab:tbad_ent_deviations}
    \centering
    \begin{tabularx}{\linewidth}{ @{\extracolsep{\fill}} *{13}c @{} }
        \midrule \midrule
        TBAD\textsubscript{EXT} & \multicolumn{2}{c}{$\Delta$SNR = 30} & \multicolumn{2}{c}{$\Delta$SNR = 10} & \multicolumn{2}{c}{$\Delta$SNR = 30} & \multicolumn{2}{c}{$\Delta$SNR = 10} & \multicolumn{2}{c}{$\Delta$SNR = 30} & \multicolumn{2}{c}{$\Delta$SNR = 10}                                                                                                 \\
        \cmidrule{2-5}\cmidrule{6-9}\cmidrule{10-13}
                                & \(e_{\Sigma}\)                       & \(e_{\max}\)                         & \(e_{\Sigma}\)                       & \(e_{\max}\)                         & \(e_{\Sigma}\)                       & \(e_{\max}\)                         & \(e_{\Sigma}\) & \(e_{\max}\) & \(e_{\Sigma}\) & \(e_{\max}\) & \(e_{\Sigma}\) & \(e_{\max}\) \\
        \cmidrule{2-3}\cmidrule{4-5}\cmidrule{6-7}\cmidrule{8-9}\cmidrule{10-11}\cmidrule{12-13}
        FL                      & 4.06\%                               & 3.48\%                               & 4.26\%                               & 3.57\%                               & 0.58\%                               & 0.50\%                               & 1.00\%         & 0.84\%       & 1.19\%         & 2.08\%       & 1.52\%         & 2.10\%       \\
        TL                      & 3.04\%                               & 3.84\%                               & 3.70\%                               & 4.15\%                               & 1.82\%                               & 2.56\%                               & 2.62\%         & 3.09\%       & 3.45\%         & 7.64\%       & 4.32\%         & 8.14\%       \\
        Outlet                  & 1.91\%                               & 1.99\%                               & 2.13\%                               & 2.26\%                               & 0.94\%                               & 1.21\%                               & 1.35\%         & 1.58\%       & 2.06\%         & 4.73\%       & 2.45\%         & 4.87\%       \\
        \midrule \midrule
        Average                 & 1.77\%                               & 1.74\%                               & 2.33\%                               & 2.16\%                               & 1.10\%                               & 1.59\%                               & 1.68\%         & 2.12\%       & 1.78\%         & 3.72\%       & 2.45\%         & 4.36\%       \\
        \bottomrule
    \end{tabularx}\label{tab:tbad_ext_deviations}
\end{table*}
\subsection{Patient Cases}
\begin{figure}
    \centering
    \includegraphics{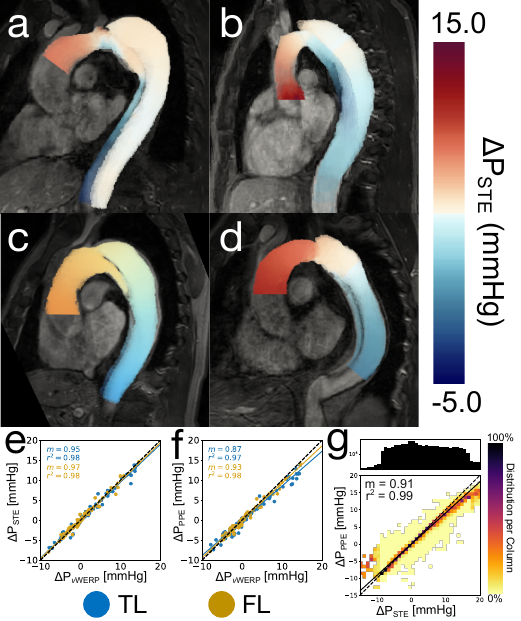}
    \caption{\textit{In vivo} analysis results. (a) STE relative pressure field of a high-growth patient case at peak systole. (b) STE relative pressure field of a low-growth patient case at peak systole. (c) STE relative pressure field of a second high-growth patient case at peak systole. (d) STE relative pressure field of a second low-growth patient case at peak systole. (e) TL and FL correlation plot of STE vs.\ \textit{v}WERP across all patient cases. (f) TL and FL correlation plot of PPE vs.\ \textit{v}WERP across all patient cases. (g) Fully spatial correlation plot of PPE vs. STE across all patient cases.}\label{fig:patient_overview_results}
\end{figure}

For all eight patient cases, PPE and STE produced relative pressure fields that were consistent with our expectations---relative pressure was highest in the aortic root and decreased when moving distally.
Pressure drops were visibly higher in the TL, where a majority of the net flow was present~\cite{xuComputedTomographybasedHemodynamic2021}.
Pressure drops in the FL were generally lower than in the TL, although the magnitude of this trend was dependent on the calculated growth rate of each patient~\cite{marleviFalseLumenPressure2021}.
Lower FL growth rate patients had higher pressure drops across the FL, and higher FL growth rate patients had lower pressure drops across the FL (\autoref{fig:patient_overview_results}abcd).
However, with $N=8$ patient cases, further study must be done to establish population-level trends between FL GR and relative pressure drops.

The PPE, STE, and \textit{v}WERP pressure drops were analyzed for agreement in \autoref{fig:patient_overview_results}efg.
Correlations between all relative pressure estimation methods were extremely strong (\(r^2\) = 0.97--0.98).
Pressure drops calculated by \textit{v}WERP tended to be the highest, followed by STE (\(m\) = 0.95--0.97 when compared to \textit{v}WERP), and finally PPE (\(m\) = 0.87--0.93 compared to \textit{v}WERP).
This ordering of the results matched the outcome of the \textit{in vitro} phantom analysis, where \textit{v}WERP also produced the highest pressure drops.

Standard practice was to run the patient cases with nearest-neighbor upsampling by a factor of two.
At this level, relative pressure estimations were completed in a matter of minutes as timed by MATLAB's \texttt{timeit} function.
Timing results for a representative patient case were: 527.4 s for STE, 151.4 s for PPE, 192.7 s for \textit{v}WERP when calculating the FL pressure drop, and 41.3 s for \textit{v}WERP when calculating the TL pressure drop. Note that STE and PPE have a single time (rather than separate FL and TL timing) because it is possible to calculate all possible drops over the domain simultaneously. If a single drop for STE or PPE were desired, it would be possible to restrict the domain to achieve a faster runtime, but this would negate the benefit of having spatially comprehensive relative pressure fields.

%
%

\section{Discussion}\label{sec:discussion}
\normalcolor
The relative pressure estimators \textit{v}WERP, STE, and PPE were evaluated across \textit{in silico}, \textit{in vitro}, and \textit{in vivo} TBAD datasets.
The \textit{in silico} analysis was performed on a CFD model with varying spatiotemporal resolution and synthetic noise levels.
Notably, the \textit{in silico} analysis allowed for comparisons between estimated relative pressure fields and the ground truth CFD solution on a voxel-by-voxel basis.
This allowed for a novel analysis of the spatial distributions of errors.
The \textit{in vitro} analysis was performed on a compliant TBAD flow phantom with three unique entry and exit tear variants.
Twelve pressure catheter measurements were made in each TBAD flow phantom tear variation, combining the benefits of acquired 4D-flow data with significantly more ground truth pressure measurements than would be feasible clinically.
The \textit{in vivo} analysis was performed on eight patient cases.
Although no ground truth was available, this analysis provided a qualitative view of relative pressure fields and an opportunity to observe the relative pressure estimators in a clinical context.

\subsection{Spatial Error Distribution}
An important novelty of the fully spatial \textit{in silico} analysis was the ability to precisely analyze the spatial variation of errors in the estimated relative pressure fields.
In general, errors were highest on the vessel wall boundaries and in areas with high velocity gradients.
These two factors combined to become particularly impactful in the 3.0 mm \textit{in silico} images---an extremely narrow region in the proximal TL had high velocities, leading to sharp velocity gradients and extensive partial voluming.
Thus, absolute errors exceeding 15 mmHg and relative errors exceeding 100\% occurred, as seen in~\autoref{fig:origin_shift}b and~\autoref{tab:relative_error_no_noise}.
To further corroborate that partial voluming/interpolation was causing high errors, an additional ``origin shift'' analysis was performed by translating the CFD model by 1/3 of a given voxel size.
The details of this analysis are summarized in Appendix A.

Because the relative pressures were calculated in reference to the TL outlet, errors in the proximal TL shifted the rest of the estimated relative pressure field by a constant for both STE and PPE\@.
This behavior leads to high errors when compared to the ground truth CFD data (and when compared to the TL pressure drop ground truth), but \textit{not} when compared to the FL pressure drop ground truth.
Because the rest of the model proximal to the narrow TL was disrupted by the same constant, the four FL pressure drops we probed remained accurate.
Errors generally increased when moving proximally from the TL outlet, which is expected, given that the relative pressure fields were shifted in reference to the TL outlet pressure.

In the patient cases, small segments of 2--8 voxels with erroneous velocity measurements led to small regions (2--8 voxels as well) with large relative pressure errors.
High (but localized) erroneous velocity gradients led to corresponding localized relative pressure errors.
In response to these findings, patient data were preprocessed to remove 4D-flow voxels whose magnitudes/orientations differed significantly from those of their neighbors.
If STE and PPE are used clinically, care must be taken to appropriately filter out erroneous velocity measurements to ensure errors are not magnified in the corresponding relative pressure fields.

\subsection{Effect of Resolution}
Increased spatial resolution generally increased the accuracy of the estimated relative pressure fields.
Accuracy decreased when temporal resolution decreased, mainly because the relative pressure estimators did not have the time frames available to capture the highly transient pressure drop at peak systole.
Errors were primarily caused by underestimation of the relative pressure extrema (both systolic peaks and diastolic valleys) and a temporal shift in the latter half of the cardiac cycle.
This behavior was more drastic in the 4D-flow MRI acquisitions compared to the synthetic 4D-flow images.
The synthetic 4D-flow images were constructed with instantaneous velocity measurements, which meant there was no temporal averaging.
Temporal averaging, of course, occurred in the 4D-flow scans.
In the \textit{in silico} analysis, the estimators may have missed peak relative pressures, but they accurately estimated the relative pressures based on the time frames they had access to.
Meanwhile, the 4D-flow scans both undersampled the peak systolic pressure drop and temporally averaged the time frames that were sampled, leading to more pronounced underestimation.
The patient scans had anisotropic voxels, like many clinical scans.
This may result in velocity gradients having varying accuracy in different coordinate directions, potentially underestimating relative pressure gradients.
Additionally, segmentation quality has been shown to affect relative pressure estimation accuracy~\cite{nolteValidation4DFlow2021}, with more conservative segmentations (inclusion of fewer boundary voxels) leading to more accurate results, particularly for PPE\@.
It is likely that this behavior would be even more pronounced in flow regimes dominated by viscous effects~\cite{lamataAorticRelativePressure2014}.
An additional analysis where the segmentations of the \textit{in vitro} flow phantoms were systematically perturbed is summarized in Appendix B.

\subsection{Effect of Noise}
The average relative errors across all \textit{in silico} noisy realizations were roughly equivalent to the cases with no added noise.
Whereas higher resolution was almost always better in the SNR = inf cases, this trend did not precisely hold in the noisy cases.
Low spatial resolution combined with high temporal resolution resulted in larger deviations in estimated relative pressures, although the average relative pressures across noise realizations were still close to the SNR = inf cases.
For example, the STE relative error ranges were larger in the 2.0 mm $\times$ 20 ms spatiotemporal resolution results compared to the 2.0 mm $\times$ 40 ms and 60 ms results when SNR = 10.
This is likely due to no smoothing filters being applied to the data to stabilize the effects of localized noise.
Deviations from the baseline (no added noise) relative pressure drops were also routinely higher in the TL compared to the FL across both \textit{in silico} and \textit{in vitro} analyses.
TBAD presents several challenging phenomena, but these results indicate that low spatial resolution relative to lumen diameter is more problematic than the relatively lower SNR present in the FL.
All relative pressure estimators were more sensitive to noise in the \textit{in vitro} analysis compared to the \textit{in silico} analysis.
Some degree of noise was already present in the \textit{in vitro} scans, though this was not substantial.
The only difference in the noise-addition methodology between the \textit{in silico} and \textit{in vitro} analyses---the usage of an acquired magnitude image---may partially explain the difference in noise performance \textit{in vitro}.
This suggests that future synthetic analyses should use more advanced magnitude-generation techniques to more accurately model clinical scenarios.

Interestingly, there was only a minor decrease in performance when increasing the noise level from $\Delta$SNR = 30 to $\Delta$SNR = 10 in the \textit{in vitro} deviation results.
The actual standard deviations of the noise fields added to the ground truth 4D-flow acquisition, calculated by finding the spatial standard deviation of $u_\text{noise}-u, v_\text{noise}-v,$ and $w_\text{noise}-w$ for each time step, were 0.025--0.028 m/s for $\Delta$SNR = 30 and 0.043--0.045 m/s for $\Delta$SNR = 10 on average for a representative TBAD\textsubscript{ENT} noise realization.
This model variation had a VENC of 1.7 m/s.
This means that the perturbation of each velocity component was on the order of a single percent of the velocity magnitude, particularly in high-velocity regions such as the ascending aorta or the TL at systole.
These are the values that were observed following the noise addition procedure outlined in~\cite{ferdian4DFlowNetSuperResolution4D2020}, which focused on the simulation of noisy 4D-flow for superresolution and denoising.
It should be noted that the signal power was based on the magnitude image, as is done in most SNR estimations.
This was done to more closely match the expected velocity corruption levels based on standard SNR measurements.
Nolte \textit{et al.}~\cite{nolteValidation4DFlow2021} took a different approach, more directly perturbing the velocity fields by specifying the standard deviation of the bivariate Gaussian noise according to the VENC and desired noise level.
While this method is less representative of how SNR is measured, it did allow for a more direct relationship between prescribed noise level and degree of velocity corruption.
Consequently, they observed much wider ranges of relative pressure drops across noise realizations in their relative pressure estimations, particularly in the higher-VENC models with smaller coarctation diameters.

\subsection{Summary of Relative Pressure Estimators and Clinical Application}
\begin{figure}
    \centering
    \includegraphics{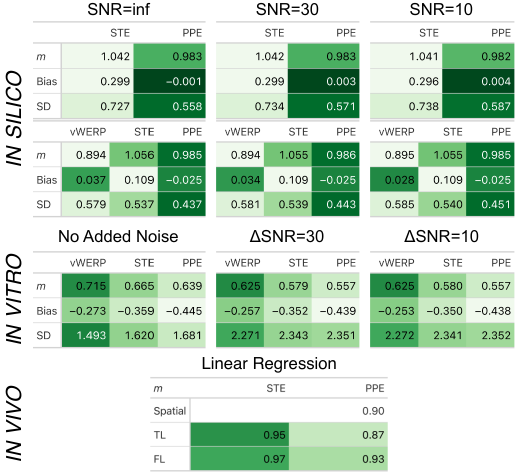}
    \caption{Overview condensing the linear regression and Bland--Altman analyses across all test beds.
        Each statistic is colored according to its relative performance per row. That is, each row of each table has a unique scale in order to make performance differences per analysis obvious.
        Each color scale is set such that the ideal outcomes ($m=1$, Bias = 0, and SD = 0) are dark green.
        The most extreme deviation from the ideal outcome is colored light green.
        The \textit{in silico} statistics were averaged across all spatiotemporal resolutions, with the 3.0 mm cases excluded as the resolution was insufficient to resolve flow.
        The \textit{in vitro} statistics were averaged across every model variation. Because the \textit{in vivo} cases had no ground truth, the TL and FL correlations are made with respect to \textit{v}WERP (as in \autoref{fig:patient_overview_results}). The fully spatial slope is made with respect to STE.}\label{fig:grand_overview}
\end{figure}

\begin{figure}[h]
    \centering
    \includegraphics{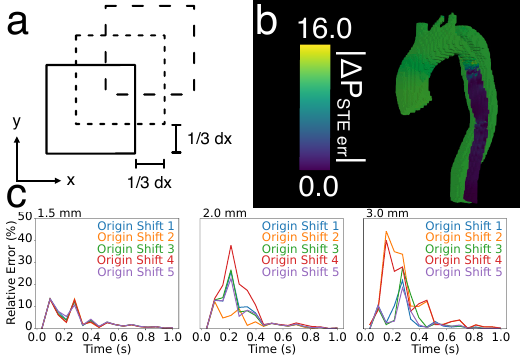}
    \caption{Overview of \textit{in silico} origin shift analysis for STE\@. (a) Overview of the origin shift method for one voxel in 2D. Each square indicates a different shift in the origin by a subvoxel amount, forcing a different interpolation\@. (b) Example of the large errors in the FL and proximal TL after an unfavorable shift\@. (c) Relative errors over the cardiac cycle for five different origin shifts. Temporal resolution was fixed to 60 ms for each variation in spatial resolution.}\label{fig:origin_shift}
\end{figure}

Each relative pressure estimator recovered flow phantom ground truth relative pressures with a minimum correlation coefficient of 0.9 and a tendency to mildly underestimate peak pressure drops (see Fig.~\ref{fig:grand_overview}).
All methods could reasonably recover ground truth relative pressures.
Based on our analysis, \textit{v}WERP outperformed STE and PPE when calculating plane-to-plane pressure drops.
It had the highest accuracy in the 4D-flow MRI acquisitions and was significantly faster than STE.
Although \textit{v}WERP performed worse than PPE and STE in the \textit{in silico} cases, \textit{v}WERP's improved performance in the \textit{in vitro} cases suggests that there are aspects of MRI simulation that we are missing in our synthetic MRI generation methodology.
When a fully 3D relative pressure field is desired, the ideal relative pressure estimator is STE\@.
Our data suggest that STE performs better than PPE at lower spatial resolutions, as its performance did not decline as substantially in the 3.0 mm \textit{in silico} cases.
Similar to~\cite{nolteValidation4DFlow2021}, STE was more accurate than PPE in the \textit{in vitro} analysis.
It should be noted that Nolte \textit{et al.}~\cite{nolteValidation4DFlow2021} used a different implementation of PPE and STE, utilizing backward differences for temporal derivatives (rather than midpoint) and an FEM\@.
Although PPE was faster than STE (roughly 2.5 minutes vs. 8.7 minutes of CPU time), both methods calculated patient relative pressure fields in a reasonable amount of time.

Clinically, relative pressure estimation could provide valuable diagnostic information for cardiovascular disease, particularly in cases of hemodynamic obstruction (e.g., valve stenosis, coarctation) or where pressure could provide a biomarker of disease~\cite{dushfunian4DFlowMRI2025}.
In the case of hemodynamic obstruction, noninvasive estimation~\cite{riesenkampffPressureFieldsFlowSensitive2014,saittaEvaluation4DFlow2019} has been shown to be possible, but ensuring accuracy of these techniques remains critical.
While relative pressure drops are not currently used in clinical practice for TBAD, previous work~\cite{marleviFalseLumenPressure2021} indicates that FL maximal pressure drop is predictive of future FL GR.
Others have also suggested the concept of \emph{first balance point}~\cite{xuComputedTomographybasedHemodynamic2021} (an indicator measuring the distance from the entry tear where pressures across the intimal flap equilibrate) as an indicator of further growth in TBAD.
As both aortic diameter and aortic growth rate are drivers of clinical decision-making~\cite{doriaCriticalReviewGuidelines2025}, predicting future growth using relative pressure could provide a valuable early diagnostic indicator.
Examining the utility of the methods introduced here, we note that the maximal FL pressure drops (per meter) in~\cite{marleviFalseLumenPressure2021} ranged from 21 mmHg/m to 72 mmHg/m.
The differences in slope between \textit{v}WERP, STE, and PPE in the \textit{in vitro} analysis suggest that all relative pressure estimators would be similarly predictive of clinical FL GR, though more work is required to understand the systematic underestimation of invasive measures.
While \textit{v}WERP showed modestly improved accuracy, whether further features of pressure fields provided by STE and PPE could give additional diagnostic insights (particularly around the inlet tear) requires further study.

In vascular diseases where plane-to-plane pressure gradients already anchor clinical decisions (e.g., coarctation, aortic stenosis), \textit{v}WERP is the natural candidate for translation, given its superior \textit{in vitro} accuracy for plane-to-plane pressure drops across all tested model variations and noise levels.
Still, establishing estimated pressure equivalences to catheter or Doppler-derived pressure thresholds would require dedicated validation for each disease.
It should be noted that disease-specific validation by pressure catheter would require measurement of \textit{instantaneous} relative pressure drops---not peak-to-peak, as used in coarctation guidelines---since relative pressure estimators do not recover the absolute pressures measured by a pullback procedure.

Both STE and PPE constitute novel methods of characterizing cardiovascular function, and no clinical guidelines currently utilize relative pressure \emph{fields}.
STE is well suited to the complex and highly variable anatomies found in TBAD and repaired type A aortic dissection (TAAD), since multiple characterizations of vascular function can be made from a single computed pressure field: interluminal pressure differences (first balance point, malperfusion/obstruction risk) and pressure drops in a single lumen (false lumen maximal pressure drop).

\subsection{Limitations and Future Work}
A major need for clinical translation is the development of deep-learning segmentation methods for TBAD in MRA/MRI~\cite{mastrodicasaArtificialIntelligenceApplications2022}.
Our study used semi-automated segmentation methods, which provided accurate research-grade results at the expense of manual effort.
Several practical pathways exist toward DL-compatible pipelines for TBAD\@.
MRA in regions of well-developed flow provides strong vessel-lumen contrast and is amenable to CNN-based segmentation of the kind already demonstrated for non-dissected aortas, though dissection-specific difficulties---intimal flap delineation and mural thrombus in the false lumen---remain challenging problems.
When MRA is acquired in the same session as 4D-flow MRI, registration burden is minimal, enabling mapping of MRA-derived segmentations into the 4D-flow domain.
Alternatively, contrast-enhanced or multi-venc acquisitions could improve signal in the patent false lumen and enable training directly on 4D-flow, although characterization of mural thrombus would likely require complementary tissue-contrast sequences.
Future work should address both the development of DL segmentation methods for TBAD and the quantification of their impact on relative pressure estimation accuracy.
Beyond segmentation, deep learning may also improve the fidelity of the underlying data---recent work has explored superresolution in both space~\cite{ericssonGeneralizedSuperResolution4D2024} and time~\cite{callmerDeepLearningTemporal2026}, and the impact of these methods on pressure estimation warrants investigation, particularly given the tendency for peak pressures to be underestimated in acquired 4D-flow (\textit{in vitro}, \textit{in vivo}).

The synthetic MRI generation protocol we used~\cite{ferdian4DFlowNetSuperResolution4D2020} could be improved with more advanced modeling approaches~\cite{dirixSynthesisPatientspecificMultipoint2022,weineCMRsimPythonPackage2024} to better capture hypothetical clinical scenarios.
The CFD model design and synthetic image generation procedure were both reasonably straightforward; although the model was sufficiently representative for this study, future work could incorporate FSI with compliant walls.
The current rigid wall assumption is a simplified model of aortic tissue, and we should expect observed relative pressure trends to change mildly in a more physically realistic model~\cite{zhuFluidStructureInteractionSimulations2022}.
The interpolation to a regular grid required that an extremely large number of elements (roughly 2.6 million) be present to preserve higher-order derivatives when projecting to the finer grid resolutions, hence the simpler modeling choices.
A voxel anisotropy sensitivity analysis is also warranted, as anisotropic voxels remain widely used in clinical scans in order to reduce scan time despite recommendations for isotropic voxels~\cite{bissell4DFlowCardiovascular2023}.

All analyses were performed in the context of type B aortic dissection---while this condition encompasses varied flow regimes that may generalize to other pathologies (stenosis, etc.), it is not a substitute for disease-specific assessment.
Moreover, the pressure drops assessed in TBAD are typically moderate compared to stenoses or aortic coarctations, which could bias the comparison of relative pressure estimators if hemodynamic sources driving relative pressure differ across regimes.
Finally, the \textit{in vivo} analysis did not include patient catheter data, though such invasive data are not typically acquired (particularly in the FL) in patients with aortic dissection.
Future work on the clinical translation of TBAD relative pressure drops should include \textit{in vivo} pressure catheter measurements, 4D-flow MRI acquisitions, and growth rate measurements, all in significantly larger patient cohorts.
Instantaneous pressure drops would need to be measured with a dual catheter setup as in~\cite{nolteValidation4DFlow2021}.
This will make it possible to directly quantify relative pressure estimation error \textit{in vivo}, which will be essential when defining clinically acceptable error thresholds.

%
%

\section{Conclusion}\label{sec:conclusion}
In this work, we analyzed \textit{v}WERP, STE, and PPE over \textit{in silico}, \textit{in vitro}, and \textit{in vivo} type B aortic dissection datasets.
\autoref{fig:grand_overview} gives a quantitative summary of these analyses.
In the fully spatial \textit{in silico} analysis, PPE was shown to outperform STE in all metrics, showing less bias ($4\times10^{-3}$ vs. 0.30), less deviation (0.60 vs. 0.74), and a slope closer to unity (0.98 vs. 1.04).
In the \textit{in vitro} analysis using plane-to-plane measurements compared to catheter recordings, \textit{v}WERP moderately outperformed both STE and PPE in accuracy (with slope of 0.72 vs. 0.67 and 0.64, respectively), bias (-0.27 vs. -0.36 and -0.45), and deviation (1.50 vs. 1.62 and 1.68).
Finally, the relative pressure estimators were evaluated for agreement over patient cases as a clinical feasibility assessment, as no ground truth pressure catheter data were available.
PPE and STE agreed strongly, and \textit{v}WERP tended to estimate the largest pressure drop, mimicking the trend of the \textit{in vitro} analysis.
All relative pressure estimators performed well in the tested aortic dissection cases, demonstrating resilience to noise and ability to estimate the ground truth relative pressure, albeit with a tendency to underestimate peak relative pressures.
The time to compute the relative pressure estimations was on the order of minutes, demonstrating their feasibility for integration into a clinical workflow.
While there was no clear winner across \emph{all} levels of analysis (\textit{in silico, in vitro, in vivo}), excluding the \textit{in silico} results gave a clear ordering: \textit{v}WERP was the best across all metrics, then STE, then PPE.
However, if a fully spatial pressure field is desired, STE should be the estimator of choice.
Further work is needed to determine why relative performance varies according to noise modeling, particularly to understand the performance decline from \textit{in silico} to \textit{in vitro}.

%
%
\section*{Appendix A: Origin Shift Analysis}
Here the origin shift analysis is summarized.
This was an auxiliary analysis done to further corroborate that partial voluming/interpolation was causing high errors in the 3.0 mm \textit{in silico} analysis.
This was done by translating the CFD model by 1/3 of a given voxel size, i.e., 0.5 mm, 0.67 mm, and 1.0 mm.
Each translation forced different interpolations of the boundary velocities from the CFD ground truth, changing TL pressure drops and relative error.
One can imagine a 6 mm diameter region that is accurately represented by two 3 mm voxels.
A 1 mm translation would then cause the same data to be represented over three voxels with partial voluming.
In the 1.5 mm and 2.0 mm cases, the TL pressure drop was resilient to shifting the model due to the increased resolution relative to the TL thickness; however, the 3.0 mm case gave drastically different pressure drops depending on the shift, ranging from 10.9 mmHg to 18.2 mmHg as the maximum pressure drop.

\section*{Appendix B: Segmentation Perturbation Analysis}
\begin{figure}
    \centering
    \includegraphics{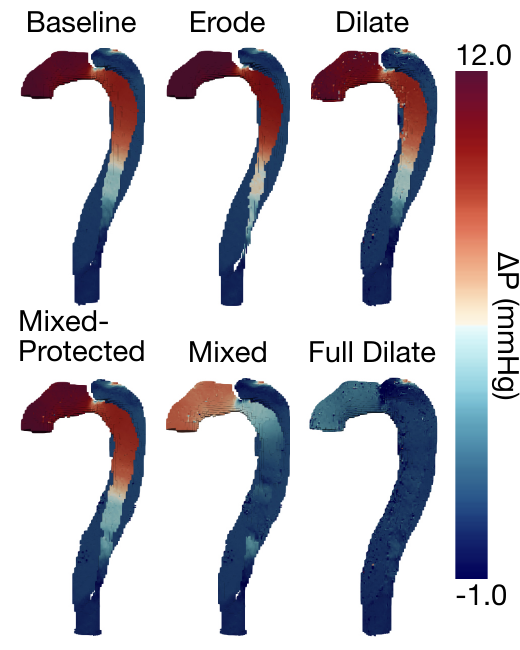}
    \caption{STE-derived relative pressure fields for TBAD\textsubscript{ENT} at peak systole, shown for each segmentation variant. The original segmentation is shown in the top left, and the segmentation variations are shown in the other five panels. The full dilate case, which completely ignored the intimal flap, produced the worst results. The mixed case without flap protection produced the second worst results. The erode, dilate, and mixed-protected cases were substantially more similar to the baseline case. Note that the random erosions and dilations made on the mixed and mixed-protected cases were made by converting the boundary voxels to a linear index, selecting evenly spaced indices, and mapping back to the original domain.}\label{fig:reproducibility}
\end{figure}

An additional analysis was performed to evaluate the sensitivity of the STE- and PPE-derived relative pressure fields to systematic perturbations of their segmentations.
Five segmentation perturbations were applied to each \textit{in vitro} model for both STE and PPE; \textit{v}WERP was not included here, as its segmentation sensitivity had already been evaluated in Marlevi \textit{et al.}~\cite{marleviFalseLumenPressure2021}.
The variants were defined as follows: a single-layer erosion of the original segmentation (erode), a single-layer dilation with the intimal flap preserved (dilate), a random mixture of erosions and dilations within a 5-voxel boundary band without preserving the flap (mixed), the same random mixture with the flap preserved (mixed-protected), and a single-layer dilation without preserving the flap (full dilate).
These five variants were applied across all three \textit{in vitro} models to examine relative pressure estimates from both STE and PPE, yielding 30 separate estimations.
Each combination of model, method, and segmentation was compared to the baseline relative pressure field by calculating the total deviation according to~\eqref{eq:total_rep_deviation}.
\begin{equation}\label{eq:total_rep_deviation}
    \text{Total Deviation}\ = \ \frac{\lVert \hat{p}_{\text{perturbation}} - \hat{p}_{\text{baseline}}\rVert_{0,2}}{\lVert \hat{p}_{\text{baseline}} \rVert_{0,2}}
\end{equation}
Linear regression and Bland--Altman plots were also generated to assess voxel-wise agreement between the perturbed and baseline relative pressure fields.
Results are summarized in~\autoref{tab:reproducibility_ranges}, with ranges reported across the three \textit{in vitro} models for each estimator and segmentation perturbation type.
\autoref{fig:reproducibility} shows the relative pressure fields for the STE/TBAD\textsubscript{ENT} combination across each segmentation variant at peak systole.
STE was more resilient to changes in the segmentation mask than PPE, consistent with the segmentation sensitivity findings reported by Nolte \textit{et al.}~\cite{nolteValidation4DFlow2021}.
Among the variants, full dilation---which ignored the intimal flap---consistently produced the worst agreement with baseline.
The unprotected mixed case produced the second-worst results: in each instance, 1--2 spurious connections were introduced between the TL and FL.
The erode, dilate, and mixed-protected variants remained substantially closer to baseline, indicating that preventing spurious connections between the lumina was the dominant factor governing segmentation sensitivity.
Erosion introduced a mild positive bias and dilation a mild negative bias.
A ``speckling'' effect was visible on the surface of the dilated cases, attributable to the inclusion of static voxels and the resulting corrupted gradients.
Finally, deviations in the erode cases were largest in already-narrow regions: the distal TL, the entry and exit tears, and a small region immediately distal to each tear.

\begin{table*}
    \centering
    \caption{Deviation Metrics for Segmentation Perturbation Analysis}
    \begin{threeparttable}
        \begin{tabularx}{\linewidth}{ @{\extracolsep{\fill}} *{6}c @{} }\toprule
            STE             & Total Deviation (\%) & \(m\)      & \(r^2\)    & Bias                   & Standard Deviation \\
            \cmidrule{2-6}
            Erode           & 2.53--9.29           & 0.99--1.07 & 1.00--1.00 & \(0.00\) to \(0.09\)   & 0.08--0.25         \\
            Dilate          & 3.51--5.36           & 0.97--0.98 & 1.00--1.00 & \(-0.09\) to \(-0.05\) & 0.07--0.10         \\
            Mixed-Protected & 3.85--5.18           & 0.99--1.01 & 1.00--1.00 & \(-0.04\) to \(0.01\)  & 0.05--0.20         \\
            Mixed           & 15.3--42.3           & 0.64--0.89 & 0.90--0.97 & \(-0.48\) to \(-0.09\) & 0.27--1.10         \\
            Full Dilate     & 52.2--85.2           & 0.21--0.59 & 0.34--0.80 & \(-2.05\) to \(-0.33\) & 0.76--2.23         \\
            \midrule\midrule
            PPE             & Total Deviation (\%) & \(m\)      & \(r^2\)    & Bias                   & Standard Deviation \\
            \cmidrule{2-6}
            Erode           & 4.83--21.4           & 1.00--1.17 & 0.99--1.00 & \(0.07\) to \(0.20\)   & 0.15--0.55         \\
            Dilate          & 7.40--12.8           & 0.92--0.95 & 0.99--1.00 & \(-0.19\) to \(-0.11\) & 0.15--0.21         \\
            Mixed-Protected & 2.27--5.16           & 0.98--0.99 & 1.00--1.00 & \(-0.02\)              & 0.06--0.08         \\
            Mixed           & 32.0--38.1           & 0.68--0.79 & 0.91--0.96 & \(-0.73\) to \(-0.19\) & 0.45--0.98         \\
            Full Dilate     & 56.9--82.2           & 0.23--0.56 & 0.41--0.76 & \(-1.84\) to \(-0.37\) & 0.80--2.09         \\
            \bottomrule
        \end{tabularx}\label{tab:reproducibility_ranges}
        \begin{tablenotes}
            \footnotesize
            \item Each table entry reports the range over each \textit{in vitro} model (TBAD\textsubscript{OR}, TBAD\textsubscript{ENT}, and TBAD\textsubscript{EXT}) for a particular combination of relative pressure estimator, segmentation type, and perturbation metric.
        \end{tablenotes}
    \end{threeparttable}
\end{table*}

\bibliography{PressureMappinginDissection.bib}
\end{document}